\PassOptionsToPackage{unicode}{hyperref}
\PassOptionsToPackage{hyphens}{url}
\documentclass[
]{article}
\usepackage{amsmath,amssymb}
\usepackage{lmodern}
\usepackage{iftex}
\ifPDFTeX
  \usepackage[T1]{fontenc}
  \usepackage[utf8]{inputenc}
  \usepackage{textcomp} 
\else 
  \usepackage{unicode-math}
  \defaultfontfeatures{Scale=MatchLowercase}
  \defaultfontfeatures[\rmfamily]{Ligatures=TeX,Scale=1}
\fi
\IfFileExists{upquote.sty}{\usepackage{upquote}}{}
\IfFileExists{microtype.sty}{
  \usepackage[]{microtype}
  \UseMicrotypeSet[protrusion]{basicmath} 
}{}
\makeatletter
\@ifundefined{KOMAClassName}{
  \IfFileExists{parskip.sty}{%
    \usepackage{parskip}
  }{
    \setlength{\parindent}{0pt}
    \setlength{\parskip}{6pt plus 2pt minus 1pt}}
}{
  \KOMAoptions{parskip=half}}
\makeatother
\usepackage{xcolor}
\usepackage{longtable,booktabs,array}
\usepackage{calc} 
\usepackage{etoolbox}
\makeatletter
\patchcmd\longtable{\par}{\if@noskipsec\mbox{}\fi\par}{}{}
\makeatother
\IfFileExists{footnotehyper.sty}{\usepackage{footnotehyper}}{\usepackage{footnote}}
\makesavenoteenv{longtable}
\usepackage{graphicx}
\makeatletter
\def\maxwidth{\ifdim\Gin@nat@width>\linewidth\linewidth\else\Gin@nat@width\fi}
\def\maxheight{\ifdim\Gin@nat@height>\textheight\textheight\else\Gin@nat@height\fi}
\makeatother
\setkeys{Gin}{width=\maxwidth,height=\maxheight,keepaspectratio}
\makeatletter
\def\fps@figure{htbp}
\makeatother
\ifLuaTeX
  \usepackage{selnolig}  
\fi
\IfFileExists{bookmark.sty}{\usepackage{bookmark}}{\usepackage{hyperref}}
\IfFileExists{xurl.sty}{\usepackage{xurl}}{} 
\hypersetup{
  hidelinks,
  pdfcreator={LaTeX via pandoc}}

\author{}
\date{}

\begin{document}

\begin{quote}
EnhancingDataIntegritythroughProvenance TrackinginSemanticWebFrameworks

NileshJain\\
\emph{technoNilesh@gmail.com (University of Mumbai)}
\end{quote}

\textbf{\emph{Abstract}---This paper explores the integration of
provenance tracking systems within the context of Semantic Web
technologies to enhance data integrity in diverse operational
environments. SURROUND Australia Pty Ltd demonstrates innovative
applica-tions of the PROV Data Model (PROV-DM) and its Semantic Web
variant, PROV-O, to systematically record and manage provenance
information across multiple data processing domains. By employing RDF
and Knowledge Graphs, SURROUND ad-dresses the critical challenges of
shared entity identification and provenance granularity. The paper
highlights the company's architecture for capturing comprehensive
provenance data, en-abling robust validation, traceability, and
knowledge inference. Through the examination of two projects, we
illustrate how provenance mechanisms not only improve data reliability
but also facilitate seamless integration across heterogeneous systems.
Our findings underscore the importance of sophisticated provenance
solutions in maintaining data integrity, serving as a reference for
industry peers and academics engaged in provenance research and
implementation.}

I. INTRODUCTION

Encompass Australia Pty Ltd (''Encompass'') is a little how-

ever unique innovation organization that has some expertise in

giving state of the art simulated intelligence and information

the executives items to both government and confidential area

markets. Established with the mission to change how associa-

tions make due, cycle, and influence information, Encompass

has quickly secured itself as a forerunner in the field by

offering special and high level arrangements. At the center of

Encompass' contributions lies its refined utilization of Seman-

tic Web information, an innovative methodology that separates

the organization from its rivals. Encompass solidly accepts

that the Semantic Web is the best method for safeguarding

significance after some time, empowering frameworks and

hierarchical changes without the deficiency of basic setting.

This conviction is grounded in the strong capacities of the

Semantic Web, which consider a more significant level of

adaptability, versatility, and versatility when contrasted with

customary information the executives strategies.

The offers for Encompass' clients are clear and convincing,

especially by they way they influence the Semantic Web

to address a large number of mind boggling information

challenges. These incentives include:

\begin{quote}
\emph{•} \textbf{Expressivity} \textbf{and} \textbf{Complexity:} The
expressivity of RDFS1and OWL22empowers the making of endlessly

complex yet strong information models. These systems

1https://www.w3.org/TR/rdf-diagram/\\
2https://www.w3.org/TR/owl2-outline/

give an establishment to addressing modern information structures that
can develop and adjust to the changing requirements of an association.

\emph{•} \textbf{Reuse of Existing Models:} The Semantic Web considers
the direct reuse of many existing, profoundly complex, and distributed
models (ontologies). This essentially de-creases the work expected to
assemble new informa-tion models without any preparation, while
additionally guaranteeing that industry best practices and laid out
information are integrated into the framework plan.

\emph{•} \textbf{Extensibility:} The RDF chart based information
struc-tures utilized by Encompass are intrinsically extensible. As
frameworks develop and advance, there is compelling reason need to
change the fundamental outline, making it simpler to oblige new
necessities without upsetting the current foundation.

\emph{•} \textbf{System Independence:} Semantic Web information de-signs
are framework free, empowering consistent in the engine framework
changes without influencing the uprightness of the information or the
applications that depend on it. This makes Encompass' answers especially
alluring to associations that expect changes in their IT framework over
the long haul.

\emph{•} \textbf{Bridging Siloed Applications:} By executing a Seman-tic
Web layer, Encompass empowers different interior applications, which are
frequently siloed and separated from each other, to flawlessly share
information. This establishes a more bound together and cooperative
cli-mate inside associations, where frameworks that were beforehand
inconsistent can now speak effortlessly.

\emph{•} \textbf{Cross-Hierarchical Information Sharing:} With the
uti-lization of Semantic Web advances, Encompass gives the capacity to
share information across authoritative limits without the requirement
for unique between hierarchical information contracts. This is made
conceivable by the Semantic demonstrating of all information components,
which guarantees that information can be perceived and utilized by
outside parties without requiring custom in-corporations.

\emph{•} \textbf{Data Validation:} The cutting edge limitation dialects,
like SHACL {[}1{]}, offer strong information approval abil-ities. These
dialects empower associations to uphold severe information quality
principles, guaranteeing that the information utilized across different
frameworks is precise, steady, and consistent with important rules.

\emph{•} \textbf{Advanced Thinking Capabilities:} The high level think-

ing capacities of OWL and SHACL permit Encompass'frameworks to construe
new information from existing in-formation. This is especially helpful
for applications that require dynamic in view of mind boggling,
interconnected information, as it empowers the framework to determine
new bits of knowledge and make expectations that could never have been
obvious through conventional strategies.
\end{quote}

All a vital rising advantage of these capacities is the capacity to give
complete provenance data across Encompass' various frameworks.
Provenance alludes to the set of experiences or ancestry of information,
including its starting points, changes, and how it has been utilized
over the long haul. This is a basic part of information the executives,
especially in situations where information precision, recognizability,
and responsibil-ity are central. With provenance data implanted in each
part of the framework, associations can follow and comprehend the
development of their information, guaranteeing that it tends to be
relied upon and that choices made in light of it are very much educated.

In this paper, we don't present new exploration claims, as this is a
\emph{Applications Track} paper. All things considered, we center around
the creative utilization of provenance in functional frameworks and the
sending of provenance-based arrangements that exhibit an experienced way
to deal with the utilization of provenance for true undertakings. Our
work ex-pects to give industry peers experiences into how provenance is
being applied inside the setting of Semantic Web advances and
information the board frameworks. Furthermore, this paper tries to
illuminate scholastics and analysts who are keen on understanding the
present status of provenance research as it connects with viable
execution. By exhibiting this present real-ity use of provenance, we
desire to give important contribution to the continuous appraisal of
provenance examination's effect and its future bearings.

The construction of the paper is as per the following: we will initially
give an outline of Encompass' extensive provenance frameworks and their
incorporation into our items and administrations. We will then examine
two explicit tasks that have used these provenance frameworks, showing
how they have been applied practically speaking. In doing as such, we
will feature the purposes for our decision of specific PROV-related
executions and portray how these decisions line up with our business
objectives and specialized necessities. At long last, we will
investigate regions where we accept provenance principles could be
improved to more readily address the issues of associations like
Encompass and its clients. Through this conversation, we expect to add
to the continuous discourse on provenance research and its reasonable
applications in the field of information the executives and simulated
intelligence.

\begin{quote}
II. SIMPLE PROVENANCE HYPOTHESIS, COMPLEX PRACTICE
\end{quote}

The extraction of helpful and noteworthy data from hetero-geneous or
enormous scope information settings is a basic test in current
information handling. This extraction can be acted in different ways
relying upon the idea of the information and the

\begin{quote}
setting where it is being utilized. In situations where a portion of the
information has a known construction, conventional questions can be
utilized to choose significant subsets of the information. The most
straightforward type of this is text-based looking, which includes
looking against printed happy with fluctuating levels of complexity,
contingent upon the idea of the inquiry question and the hidden
information. Further developed procedures might include the utilization
of factual techniques to recognize designs inside the information,
empowering the extraction of valuable data from apparently unstructured
or semi-organized datasets.

Encompass has embraced AI (ML) ways to deal with work with the
disclosure and connection of data from huge, complex datasets. Via
preparing frameworks to perceive and gather designs, Encompass' ML
frameworks can reveal stowed away connections and experiences inside
information that would somehow be challenging to distinguish utilizing
customary procedures. The utilization of ML upgrades the ability to
handle information and works on the general execution of the framework.
Close by these ML strategies, Encompass utilizes Semantic or Information
Diagram (KG)- based context ori-ented data to give extra layers of
importance and significance, working on the precision and profundity of
data recovery. These information diagrams are especially important for
grasp-ing the connections between various elements and can be utilized
to direct the translation of information, guaranteeing that setting is
safeguarded even as information develops over the long haul.

At times, the deduction of design from unstructured or semi-organized
information is likewise worked with through ML methods. These
methodologies empower Encompass to make significant, organized
portrayals from crude or boisterous information, making it more
straightforward to apply thinking and perform computerized
investigation. To additionally refine the nature of the information and
the models utilized, Encom-pass consolidates Human-in the know (HITL)
techniques into its tasks. These HITL exercises include human oversight
to survey, refine, and work on the preparation of the ML frame-works,
guaranteeing that they can be consistently refreshed and adjusted to
reflect changing prerequisites and further develop exactness. HITL
strategies are especially powerful in situations where computerized
frameworks are not adequate all alone, and where human ability is
expected to direct the educational experience.

The execution of these frameworks requires perplexing, crossover models
that join thinking, semantic web innova-tions, and AI to sort out and
recover data productively from enormous scope projects. These frameworks
should have the option to deal with information across different
organizations and cycles while guaranteeing that the provenance of all
information is kept in a deliberate and steady way. Provenance, in this
unique situation, alludes to the set of experiences or ancestry of
information, including where it came from, the way things were handled,
and the way in which it has been utilized over the long run. Provenance
is a critical part of information the executives, especially in
applications that require elevated
\end{quote}

degrees of trust, responsibility, and discernibility.

To record provenance methodicallly across numerous frame-works and
different information handling spaces, it is funda-mental to have a
clear cut and cognizant provenance refer-ence model, as well as a strong
specialized foundation for interpreting, conveying, and incorporating
information across frameworks. The presentation and inescapable
reception of the PROV Information Model (PROV-DM) {[}2{]} and its
Semantic Web partner, PROV-O {[}3{]}, has furnished Encompass with an
adaptable and exhaustive provenance structure that can be applied across
many situations. The PROV model has demonstrated to be adequately
adaptable for our necessities, with just minor augmentations expected to
fit it to the par-ticular prerequisites of our different frameworks. The
model is likewise sufficiently strong to help efficient use across our
ventures, guaranteeing consistency and interoperability.

In any case, the specialized execution of provenance fol-lowing and
joining isn't without its difficulties. Two essential difficulties that
we face in our work are:

\begin{quote}
1) \textbf{Shared Element Identification:} A basic test in
multi-framework information handling is guaranteeing that substances,
like individuals, reports, or different items, are accurately
distinguished across various frameworks.

As information moves between various frameworks and is handled in
different ways, it is fundamental to keep up with steady distinguishing
proof of these substances to protect the honesty of the provenance data.
This common element distinguishing proof guarantees that the provenance
records are exact and mirror the right connections between the
substances and their changes. 2) \textbf{Granularity:} Provenance data
should be recorded at a proper degree of detail to catch the vital
experiences without turning out to be excessively complicated or hard to
make due. The granularity of provenance alludes to the degree of detail
remembered for provenance records, and it is crucial for work out some
kind of harmony between catching adequate detail to help trust and
responsibility, while staying away from extreme intricacy that could
overpower clients or dial back the framework. Furthermore, there should
be components for totaling provenance data at more elevated levels for
framework or cycle level outlines.
\end{quote}

The principal challenge, shared element recognizable proof, is tended to
using the Asset Portrayal Structure (RDF), which utilizes special
Uniform Asset Identifiers (URIs) to address substances and different
articles. RDF's Open World Presump-tion (OWA) takes into consideration
information portrayal across independent RDF datasets, empowering
various frame-works to reference shared URIs and combine them. Encompass
use RDF for provenance following as well as the essential information
design for a large portion of its undertakings. By addressing project
information and its related provenance in RDF, we can guarantee that
elements are recognized reliably across numerous datasets, working with
the coordination of data from various sources. This approach is likewise
reached

\begin{quote}
out to non-RDF data sources, for example, Git-based program-ming and
information variant control, where URIs are utilized to distinguish and
reference substances.

To guarantee that provenance data is reliably coordinated across
different subsystems, we have fostered a bunch of rules and practices
that empower the consistent recognizable proof and following of items.
These rules include:\\
1) Item characters should be laid out in Information Dia- grams (KGs)
and got to through APIs.

2) Item character should be overseen inside the Informa-tion Diagrams at
whatever point HITL connections are required.

3) Handling subsystems should save and report standard article
personalities to guarantee consistency across var-ious phases of the
information lifecycle.

4) Reasonable arrangements of items ought to be overseen in specific
determination frameworks, like Git, as long as the datasets containing
them are portrayed inside the Information Diagrams (see dataset
granularity).

5) All handling reports and results should incorporate provenance data,
utilizing the sanctioned PROV-DM model.

6) Handling components should be rationally distinguished inside
Information Diagrams to guarantee that all parts of the framework are
precisely followed.

By keeping these rules, we have had the option to foster a strong
framework for coordinating provenance data across different subsystems
and guaranteeing that it is reliably fol-lowed and detailed all through
the information lifecycle. These frameworks and techniques are portrayed
in more detail in the accompanying segment.
\end{quote}

To delineate the adequacy of our provenance following methodology, we
give Figure 1, which shows the UI of the \emph{SURROUND Metaphysics
Platform} (SOP). This point of interaction shows provenance data inside
a Sankey chart, permitting clients to envision the progression of
information and its related provenance across various handling steps.
The provenance information showed in the chart is created by the
PROV-DM/PROV-O model by Encompass' handling work process device,
\emph{ProvWF}, which performs Named Sub-stance Acknowledgment (NER)
against electronic records and matches elements against a portion of
Encompass' Information Diagram items. As well as envisioning provenance
data, SOP additionally oversees it in packs as \emph{Managed Graphs},
which are treated as semantic resources. These Oversaw Charts are
consequently connected with provenance data, including possession and
access control, guaranteeing that information is appropriately overseen
and followed all through its lifecycle. The issue of provenance
granularity is tended to by ana-lyzing the various sorts of handling
that normally happen in heterogeneous frameworks. Encompass plays out a
for every situation evaluation of the necessary provenance granularity
for each venture, guaranteeing that the degree of detail caught is
fitting for the particular necessities of the task. Table I gives a
rundown of handling capabilities, instances of these capabil-ities, and
the necessary granularity for every situation. All by

playing out these evaluations and using existing apparatuses to create
and store provenance, Encompass can guarantee that the right degree of
provenance data is caught and kept up with for its ventures.

\begin{quote}
III. COMPANY-WIDE PROVENANCE ARCHITECTURE
\end{quote}

To productively oversee and follow the provenance of different resources
inside our IT projects, Encompass has executed an exhaustive, broad
design for recording and us-ing provenance data. This framework
guarantees that all information and cycles are detectable, irrefutable,
and can be reliably connected to their starting points, changes, and
results. Provenance following is urgent for keeping up with
straightforwardness as well as for working on the general proficiency
and trustworthiness of our information handling pipelines.

The provenance engineering we use is based on a blend of devoted
instruments and universally useful frameworks, intended to catch and
store provenance information in an or-ganized and normalized way. The
center of this engineering is the utilization of PROV-O, a broadly
embraced cosmology for demonstrating provenance in the Semantic Web.
Underneath, we depict the significant parts of this design and the jobs
they play in supporting our different information the executives needs.

\emph{A. Provenance Tools}

Our framework is fundamentally based on the accompany-ing significant
devices, each filling a particular need in the provenance following
cycle:

\begin{quote}
\emph{•} \textbf{SURROUND Metaphysics Stage (SOP)} The Encom-pass
Metaphysics Stage (SOP) is a key endeavor level information the
executives framework based on seman-tic innovations. SOP depends on Top
Quadrant's \emph{EDG} (Undertaking Information Administration) system,
which gives a vigorous establishment to overseeing information resources
and administration strategies. SOP broadens this structure by
integrating the administration of seman-tic resource states and
assortments, empowering an addi-tional adaptable and extensive
information the executives climate. SOP assumes a crucial part in
recording PROV-DM-consistent provenance for all activities including
semantic resources. This incorporates recording activi-ties performed on
semantic information resources, like changes, updates, and changes, as
well as the connections between these resources. The provenance data put
away in SOP is basic for figuring out the progression of information
across frameworks and for guaranteeing that all information changes are
straightforward and recog-nizable. Furthermore, SOP's combination with
different devices in the biological system considers consistent exchange
and representation of provenance information, adding to a bound together
way to deal with informa-tion administration across the association. SOP
records and coordinates provenance data created by different frameworks
and work processes, giving a comprehensive

perspective on the information lifecycle. For more data about SOP, visit
https://surroundaustralia.com/sop.

\emph{•} \textbf{ProvWorkflow (ProvWF)} ProvWorkflow (ProvWF) is a
Python-based structure intended to work with the making of work
processes for different information handling un-dertakings. These work
processes, once executed, record PROV-DM provenance information for the
activities per-formed during the work process execution, as well as the
information that is consumed and created. ProvWF is upheld by Encompass'
\emph{Block Library}, which gives reusable capability impedes that can
be coordinated into work processes. These blocks cover a great many
errands, for example, Information Diagram (KG) Programming interface
demands, Regular Language Handling (NLP) for text examination, and
different information handling exercises. ProvWF's measured plan
empowers the simple sythesis of intricate work processes from
straightforward, reusable structure blocks. The provenance information
produced by ProvWF is moved to SOP as provenance packs, guaranteeing
that all activities inside work pro-cesses are completely discernible
inside the more exten-sive information the executives framework. This
joining empowers start to finish following of information and cycle
changes across various phases of the work process. ProvWF likewise gives
adaptability to follow provenance at different degrees of granularity,
contingent upon the necessities of the venture. For more data about
ProvWF, visit https://surroundaustralia.com/provwf.

\emph{•} \textbf{Block Library} The Block Library is a fundamental piece
of the ProvWF biological system, containing an inventory of predefined
\emph{Blocks}, which are basically PROV-DM Activity class objects. These
blocks address reusable capabilities or activities that can be
integrated into work processes to perform normal errands, for example,
Pro-gramming interface cooperations, text handling, and in-formation
investigation. By keeping an extensive library of blocks, Encompass
guarantees that work processes can be fabricated all the more
proficiently, with normal tasks preoccupied away into reusable parts.
The Block Library improves on work process creation, decreases overt
repet-itiveness, and guarantees consistency in the execution of normal
assignments. Each block is related with its own arrangement of
provenance information, which is followed and coordinated into the more
extensive PROV-O system.

\emph{•} \textbf{Git} Git, the disseminated rendition control framework,
is utilized broadly inside our association to deal with the forming of
resources like code, information, and documentation. It permits us to
follow the progressions made to resources over the long haul and
guarantees that every adaptation is appropriately recorded and
recogniz-able. While Git itself doesn't locally uphold PROV-O
provenance, we utilize it by recording URIs for elements oversaw inside
Git vaults and referring to them in our PROV-O information. The
utilization of URIs guarantees that we can reliably follow substances
across both Git
\end{quote}

\includegraphics[width=7.13889in,height=3.54861in]{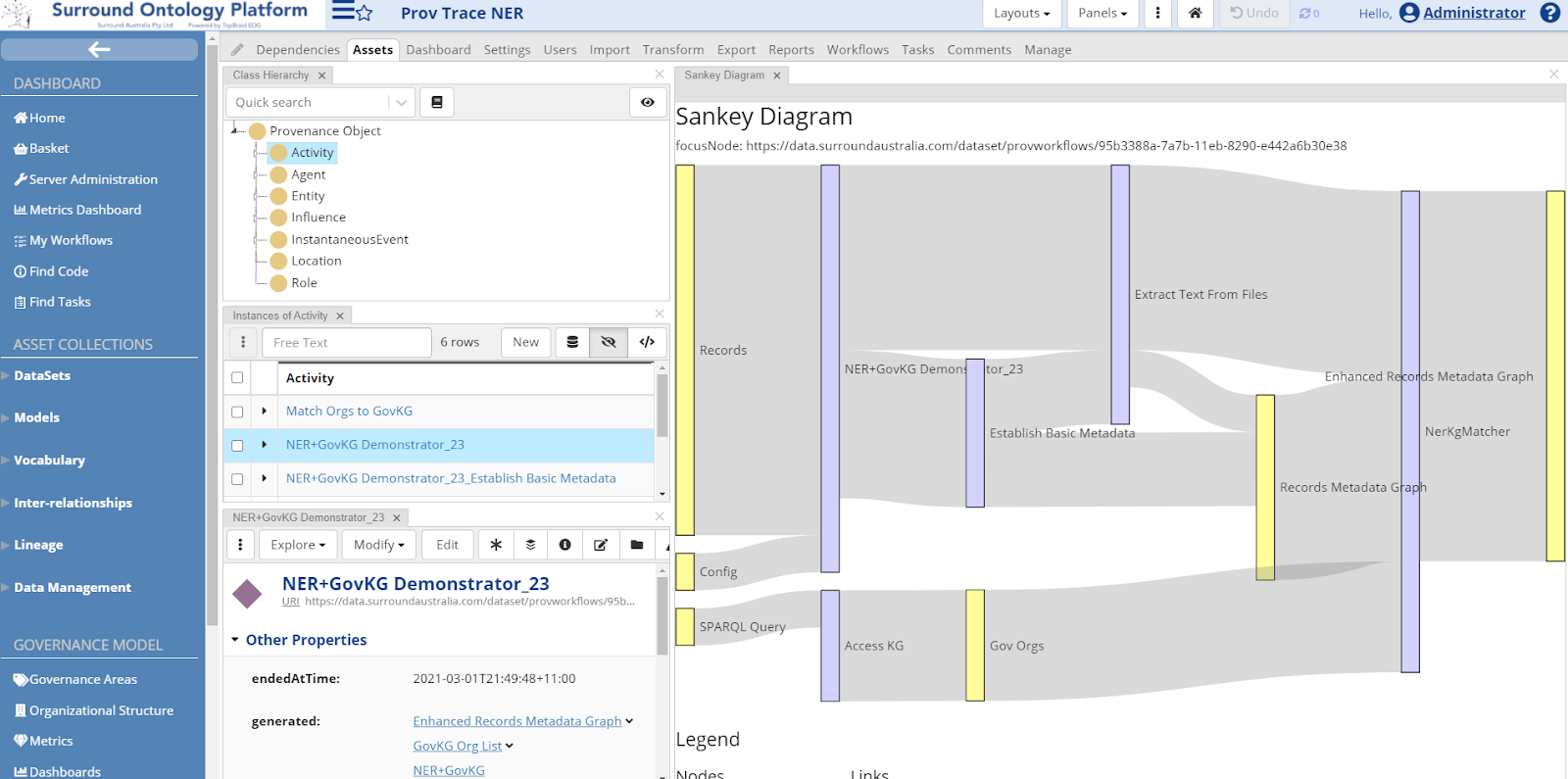}

Fig. 1. An example of a provenance trace from a processing workflow that
uses elements of a knowledge graph, performs processing in cloud-hosted
scalable services, generates augmented views of an input stream
(performing Named Entity Recognition on a document set and annotating
with elements from the knowledge graph), persists the results in the
knowledge graph and integrates the provenance trace with the provenance
trace generated by knowledge graph management.

\begin{longtable}[]{@{}
  >{\raggedright\arraybackslash}p{(\columnwidth - 4\tabcolsep) * \real{0.3333}}
  >{\raggedright\arraybackslash}p{(\columnwidth - 4\tabcolsep) * \real{0.3333}}
  >{\raggedright\arraybackslash}p{(\columnwidth - 4\tabcolsep) * \real{0.3333}}@{}}
\toprule()
\begin{minipage}[b]{\linewidth}\raggedright
\begin{quote}
\textbf{Function}
\end{quote}
\end{minipage} & \begin{minipage}[b]{\linewidth}\raggedright
\begin{quote}
\textbf{Examples}
\end{quote}
\end{minipage} & \begin{minipage}[b]{\linewidth}\raggedright
\begin{quote}
\textbf{Granularity}
\end{quote}
\end{minipage} \\
\midrule()
\endhead
\begin{minipage}[t]{\linewidth}\raggedright
\begin{quote}
Human-in-the-loop ML classification
\end{quote}
\end{minipage} & \begin{minipage}[t]{\linewidth}\raggedright
\begin{quote}
Establishment of defs, Registration entities, Annotation, Classification
for training
\end{quote}
\end{minipage} & \begin{minipage}[t]{\linewidth}\raggedright
\begin{quote}
Statement, Reified statements
\end{quote}
\end{minipage} \\
Database management, Data transformation &
\begin{minipage}[t]{\linewidth}\raggedright
\begin{quote}
Making data instances sets available in a useful form
\end{quote}
\end{minipage} & \begin{minipage}[t]{\linewidth}\raggedright
\begin{quote}
Dataset (table, spreadsheet, graph etc)
\end{quote}
\end{minipage} \\
\begin{minipage}[t]{\linewidth}\raggedright
\begin{quote}
Query
\end{quote}
\end{minipage} & \begin{minipage}[t]{\linewidth}\raggedright
\begin{quote}
Extraction of data subsets
\end{quote}
\end{minipage} & \begin{minipage}[t]{\linewidth}\raggedright
\begin{quote}
Dataset, Resultset
\end{quote}
\end{minipage} \\
\begin{minipage}[t]{\linewidth}\raggedright
\begin{quote}
Governance
\end{quote}
\end{minipage} & \begin{minipage}[t]{\linewidth}\raggedright
\begin{quote}
Selecting particular datasets for use
\end{quote}
\end{minipage} & \begin{minipage}[t]{\linewidth}\raggedright
\begin{quote}
Dataset
\end{quote}
\end{minipage} \\
\begin{minipage}[t]{\linewidth}\raggedright
\begin{quote}
Bulk object processing
\end{quote}
\end{minipage} & \begin{minipage}[t]{\linewidth}\raggedright
\begin{quote}
Indexing, classification, clustering
\end{quote}
\end{minipage} & \begin{minipage}[t]{\linewidth}\raggedright
\begin{quote}
Whole-of-workflow
\end{quote}
\end{minipage} \\
\begin{minipage}[t]{\linewidth}\raggedright
\begin{quote}
Document analysis
\end{quote}
\end{minipage} & \begin{minipage}[t]{\linewidth}\raggedright
\begin{quote}
Making information elements in a document available to finer grained
processes
\end{quote}
\end{minipage} & \begin{minipage}[t]{\linewidth}\raggedright
\begin{quote}
Document, derived dataset
\end{quote}
\end{minipage} \\
\begin{minipage}[t]{\linewidth}\raggedright
\begin{quote}
KG Management
\end{quote}
\end{minipage} & \begin{minipage}[t]{\linewidth}\raggedright
\begin{quote}
Est'ment of state of complex, modular KGs, change tracking, support for
automated up-dates
\end{quote}
\end{minipage} & \begin{minipage}[t]{\linewidth}\raggedright
\begin{quote}
Graph (Dataset)
\end{quote}
\end{minipage} \\
\bottomrule()
\end{longtable}

\begin{quote}
TABLE I\\
A LIST OF PROJECT FUNCTIONS, EXAMPLES OF THEM AND (OUR) REQUIRED
PROVENANCE GRANULARITY

stores and different information the board frameworks. This empowers
consistent incorporation of Git-oversaw resources with the more
extensive provenance environ-ment, without the requirement for complex
Git-to-PROV mappings like Git2PROV {[}4{]}. By referring to substances
in Git utilizing URIs, we can safeguard the honesty of our provenance
following and keep a reliable, brought together model of information and
cycle connections. Git storehouses can be both public and private, and
the provenance data connected with every resource is put away and
overseen in a manner that guarantees straight-forwardness and
discernibility. For more data about Git, visit https://git-scm.com/.

\emph{B. General-reason Provenance Following Tools}

Notwithstanding the significant devices referenced above, we
additionally use a few broadly useful frameworks for explicit provenance
following undertakings. These devices assist us with keeping up with
adaptability in overseeing provenance across an extensive variety of
venture types and information sources.

\emph{•} \textbf{RDFlib} RDFlib is a broadly useful Python library for
working with RDF (Asset Portrayal System) information.

It is generally utilized in our association for controlling RDF charts,
which are the central information struc-tures for addressing connections
between substances in a semantic setting. A large number of our
information objects, including provenance information, are addressed as
RDF charts, which takes into consideration reliable
\end{quote}

\includegraphics[width=7.14028in,height=2.02917in]{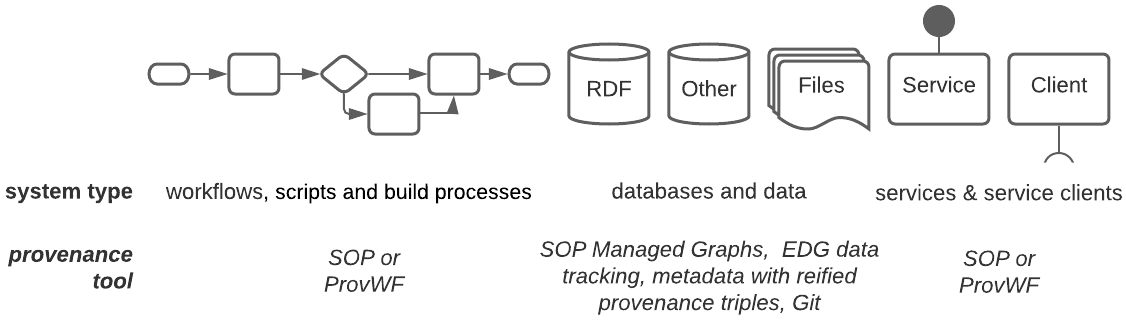}

Fig. 2. SURROUND's provenance tools linked to system type

\begin{quote}
and adaptable control. One vital component of RDFlib is its capacity to
help reified provenance, which in-cludes making definite records of the
setting in which RDF explanations are made. This reification cycle is
fundamental for keeping up with the trustworthiness of provenance data
and guaranteeing that each move made on the information can be followed
back to its starting point. We keep up with different RDFlib code blocks
to work with the creation and control of reified provenance for RDF
articulations, permitting us to protect the full history of information
changes inside our frameworks.
\end{quote}

\emph{C. Provenance Following Workflow}

The general work process for following provenance inside our association
starts with the distinguishing proof of the resources associated with a
specific venture or information handling task. Every resource is
relegated an extraordinary URI to guarantee that it tends to be
dependably referred to across various frameworks and instruments. As the
resource goes through changes --- like alterations, handling, or
exam-ination --- the provenance of each activity is kept in PROV-DM
design, catching subtleties, for example, the substance in question, the
activity performed, and the hour of the activity.

This provenance information is then incorporated into the applicable
devices and frameworks, guaranteeing that it is available for later
recovery, examination, and check. Whether through SOP, ProvWF, or Git,
the provenance information is put away in a manner that permits it to be
effectively questioned and imagined. On account of complicated work
processes, provenance packs are moved between frameworks to guarantee
that all means in the process are kept in a rational way.

By taking on this design, we guarantee that each move toward our
information handling pipelines is discernible, straightforward, and
certain, which is fundamental for keep-ing up with exclusive
requirements of information honesty, administration, and responsibility.

\begin{quote}
IV. ASPECTS OF OUR PROVENANCE MODELLING

A lot of our provenance displaying will be recognizable to PROV clients:
chains of \emph{Activities} and \emph{Entities} related with
\emph{Agents}. Notwithstanding, we have experienced a few task explicit
situations that require slight specializations. These situations, which
are definite in the accompanying venture con-textual analyses, show the
assorted utilizations of provenance in our work and the customizations
we have made to help the fluctuating necessities of various work
processes.

\emph{A. Project 1: Electronic Records Assessment}
\end{quote}

A new undertaking inside our association zeroed in on the use of Regular
Language Handling (NLP) and Information Charts (KGs) to order electronic
records for the end goal of filing. This undertaking included
extricating components of records' substance, looking at them against
oversaw well-springs of \emph{context} introduced as KGs, and utilizing
AI (ML) methods to get familiar with the ideal grouping procedures. The
test of effectively arranging such records requires nitty gritty
provenance following at different levels of the informa-tion lifecycle,
from content extraction to the use of AI models. The focal part of this
undertaking's provenance catch was the utilization of information
administrations --- our char-acterization work process framework
questioning our own KG administrations. We followed both the general
work process provenance and the singular information proclamation
provenance. The previous gave a significant level perspective on the
whole characterization process, while the last option empowered us to
approve the orders made inside a record's metadata.

\begin{quote}
Figure \textbf{??} outlines our model for following information
administration inquiries inside the work process. The full work process
provenance is crucial for follow the designs that lead to explicit
outcomes, giving straightforwardness into how various information setups
impact the last results. Every execution occurrence of the work process
and its constituent activities are recorded as PROV Activity occasions,
while the infor-mation --- like records' substance and arranged metadata
---is caught as PROV Entity cases. This permits us to keep an
\end{quote}

\includegraphics[width=7.14028in,height=1.66389in]{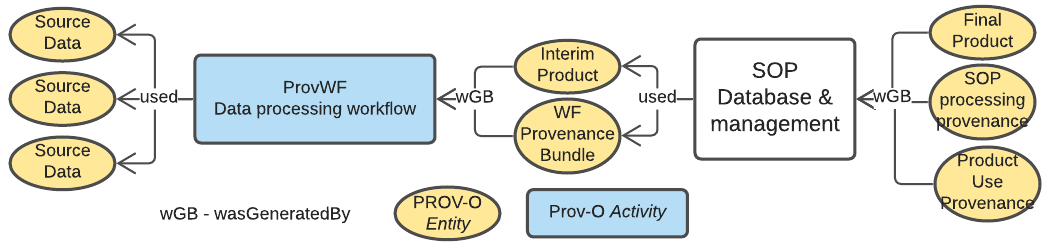}

Fig. 3. \emph{ProvWF} is often used to generate RDF data - here the
``Interim Product'' - which can be supplied to \emph{SOP} with an
acompanying provenance Bundle.

\emph{SOP}, in turn, generates both Bundles of provenance for any
actions on data it performs and also records usage provenance for
products

exact and finish record of how every information component was changed
in the interim.

Individual information explanation provenance is especially urgent for
the groupings performed on record metadata, put away in RDF design. This
permits every characterization result to be confirmed freely. Work
process provenance, metadata, and metadata provenance are completely put
away in a RDF data set, with cross-references between the components to
guarantee that all parts of the cycle are detectable and
un-questionable.

In the years since the distribution of PROV, we have noticed different
expansions and specializations custom-made for work process provenance.
One such expansion is PROV-Wf {[}5{]}, which we embraced to deal with
work process explicit provenance. Notwithstanding, we have tracked down
that the Plan, or directions for executing work processes, are normally
implanted inside the work process' characterizing programming code.
Thus, our \emph{ProvWF} instrument records a URI reference to the
particular rendition of the code (either a Git \emph{commit} or
\emph{release}) that was executed for every work process. This
guarantees that the exact form of the product that delivered the
outcomes can be followed back, keeping an unmistakable connection
between the work process execution and the basic code.

\emph{ProvWF} requires custom PROV-style logging to be char-acterized
for every custom \emph{Block} (work process part). Be that as it may, if
predefined \emph{Blocks} from our \emph{Block Library} are utilized,
this errand becomes less complex. While we have investigated frameworks
that create PROV-viable work process provenance all the more
consequently ---, for example, the methodology introduced in {[}6{]} ---
we have found that the degree of work process determination expected by
these frameworks, especially the utilization of particular business
process demonstrating dialects, surpasses the granularity of provenance
we want for our work processes. These strate-gies likewise add intricacy
to characterizing non-executable information structures, making it more
bulky than essentially characterizing PROV straightforwardly.

In late work on logical work process provenance {[}7{]}, there has been
an emphasis on demonstrating control stream to respond to questions
like: ''What are the explanations behind

\begin{quote}
different outcomes in two executions of a work process?''While we right
now don't carry out such control-stream displaying in \emph{ProvWF} or
different devices, we address work processes as a straightforward
\emph{Workflow} containing \emph{Blocks} organized directly over the
long run. All control stream choices are subsumed into the
\emph{Blocks}, which, while making them more perplexing, have permitted
us to successfully demonstrate all the significant control stream
choices inside the work process. Would it be advisable for us we choose
to zero in more intently on control stream components in later
activities, we expect to address these as specific \emph{Blocks} with
templated (anticipated) data sources and results. By contrasting
examples of these particular \emph{Blocks}, we can acquire understanding
into the particular control stream decisions made during work process
execution.

\emph{B. Project 2: Report Semantic Querying}

In another new venture, we decayed an enormous industry detail report
into primary components as well as semantic parts, for example, state
implications, equivalents, outline portrayals, phrasing records, and
calculation components. This deterioration was finished to work with
further developed normal language inquiries and to help gullible looking
through inside the report. To follow the viability of various substance
decay strategies and reference datasets --- like vocabular-ies of
industry-explicit terms --- we carried out a multi-framework provenance
following system. This permitted us to exhaustively display the
advancement of datasets and the collaborations among inquiries and their
outcomes.

We demonstrated the different datasets, large numbers of which were KGs,
as PROV \emph{Entity} examples. We followed the condition of these
datasets after some time as inquiries were made to the multi-part
framework. Each question was treated as a PROV Activity performed by an
unknown Agent, and we utilized web logs to remove results and track
changes in the KG state, similar as the strategy utilized by a portion
of our creators in past work {[}8{]}. This permitted us to catch the
provenance of question execution, including the particular datasets
questioned and the outcomes returned, giving definite bits of knowledge
into the cooperations among clients and the framework.
\end{quote}

Following the reference dataset state in this task was fin-ished
utilizing our SOP apparatus, which has consolidated chart state
following abilities throughout the long term. This empowers us to catch
provenance connected with changes, new information additions, and
different alterations inside datasets. In SOP, we can allude to the
condition of a whole assortment of resources utilizing a solitary URI
reference ---a \emph{version} of a resource assortment. This formed
reference is then utilized in the provenance records of work processes
and questions, guaranteeing that cross-questioning provenance is
conceivable. By connecting the questions and work processes to explicit
adaptations of dataset assortments, we guarantee that every one of the
information engaged with the interaction is precisely followed and
connected to its state at the hour of purpose.

This undertaking likewise elaborate utilizing the SOP appa-ratus'
ability to follow individual components inside datasets, guaranteeing
that the in general dataset as well as the parts inside it were
precisely followed after some time. This degree of granularity is
fundamental for keeping up with the honesty of the information and
guaranteeing that changes to explicit terms or definitions inside the
archive are caught as a compo-nent of the general provenance.

Through these two tasks, we have shown the adaptability and versatility
of our provenance design. Whether managing complex work processes for
record grouping or following semantic deterioration and questioning
inside an enormous report, we have utilized tweaked ways to deal with
guarantee that provenance is caught precisely and such that upholds
confirmation, straightforwardness, and reproducibility.

\begin{quote}
V. REFLECTIONS ON PROV MODELLING
\end{quote}

Throughout the span of our work with provenance dis-playing, especially
with the PROV-DM model in its PROV-O structure, we have come to
profoundly see the value in the diagram based nature of the model. The
capacity to address provenance as a diagram offers us an instinctive and
adaptable method for displaying complex connections between different
components, like \emph{Entities}, \emph{Activities}, and \emph{Agents}.
We would say, the chart based design of PROV has been key for catching
the perplexing interconnections inside multi-framework work processes,
giving an unmistakable and strong perspective on the provenance of
information and cycles.

One of the vital qualities of PROV, especially with regards to the RDF
execution, is its utilization of item distinguishing proof. By utilizing
exceptional identifiers for every element, movement, and specialist, we
can make a clear cut and tireless record of provenance across various
frameworks. This has empowered us to store provenance information in
different sorts of frameworks while as yet keeping up with the ca-pacity
to cross-question and examine this information. This adaptability is
critical, as it permits us to work with various sorts of data sets and
information stockpiling frameworks, from straightforward social data
sets to more complex chart based frameworks, without losing the
detectability of the information.

\begin{quote}
Furthermore, PROV's utilization of extensible diagrams has been
exceptionally gainful for our work. As our tasks fre-quently require
displaying complex frameworks with changing degrees of detail, the
capacity to broaden the PROV model with custom credits and connections
has permitted us to catch the essential intricacy while keeping up with
consistency and interoperability across various devices and frameworks.
The extensibility of PROV has permitted us to consolidate space explicit
data without compromising the center construction of the provenance
model, which is a critical consider guarantee-ing that the model
remaining parts both versatile and versatile.

In view of the qualities of PROV, we have had the option to show
''anything'' in our frameworks at different degrees of granularity. This
flexibility has been particularly valuable when we want to catch
provenance at various scales, from following fine-grained insights
concerning individual informa-tion components to seeing significant
level work processes and their general effects. We can store provenance
data for whole work processes, for example, the means engaged with
handling information, or for individual information components, like the
changes or choices that lead to explicit results. This capacity to show
both full scale and miniature degrees of provenance has been significant
in our capacity to perform point by point reviews, total outcomes, and
uncover bits of knowledge from complex frameworks.

In spite of the many benefits of PROV, there have been a few
difficulties and restrictions that we have experienced. These
difficulties essentially originate from situations where the implicit
abilities of PROV don't completely line up with the particular
requirements of our utilization cases. While these issues have not kept
us from effectively executing provenance models, they have expected us
to foster custom arrangements and expansions. Underneath, we dive into
the absolute most huge issues we have confronted:

1) \textbf{Difficulty in Putting away Complex Information in}
\textbf{Provenance Graphs}

\emph{•} A striking test we have experienced includes putting away
complex information objects inside our prove-nance charts. While PROV
succeeds at addressing the connections between elements, exercises, and
specialists, we frequently need to relate compli-cated, organized
information with explicit prove-nance records. By and large, we would
rather not store these perplexing articles independently from the
provenance information charts, as this would subvert the trustworthiness
of the information model and bring pointless intricacy into our
frameworks. Be that as it may, addressing complex articles inside the
provenance chart has demonstrated to be trou-blesome without depending
on excessively convo-luted Semantic Web displaying. The test lies in how
to encode these items in a way that is both proficient and simple to
oversee while safeguarding the con-nections between the various
information parts. Fur-thermore, the need to keep up with these
perplexing
\end{quote}

\includegraphics[width=7.14028in,height=2.15555in]{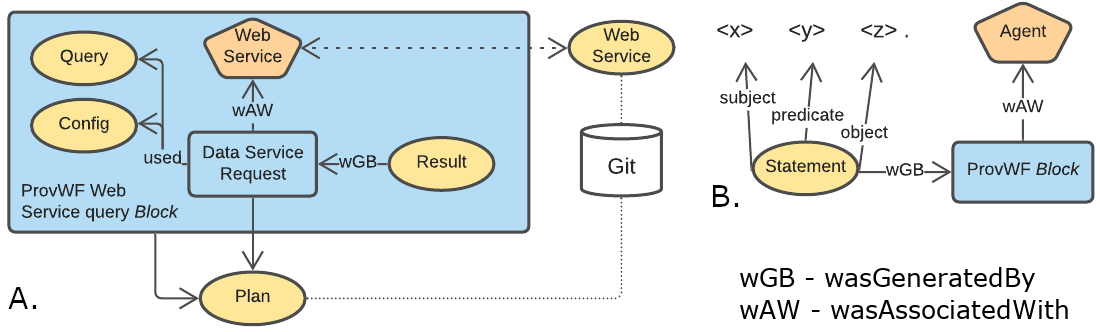}

Fig. 4. \textbf{A}. A \emph{Service query block} from our \emph{Block
Library}, implemented within our \emph{ProvWF} framework using a Query
and other configuration (Config) to

query a Web Service agent for a Result. Provenance for the web service
itself, now considered an entity, is recorded in Git systems and
referenced by each

\emph{ProvWF} execution. \textbf{B}. Reified provenance for a single RDF
triple associated with the \emph{ProvWF} Block instance that generated
it.

\begin{quote}
items inside a similar framework presents expected execution and
versatility issues, especially as the size and intricacy of the
information increment.

2) \textbf{Linking Entity Occasions to Plan Instances}

\emph{•} Another trouble we have confronted is connecting PROV Entity
occurrences to Plan examples. While PROV gives a direct method for
displaying elements and exercises, the model doesn't expressly uphold a
steady connection among Entity and Plan occurrences, which is critical
for following the work processes that created the substances. For
instance, in a portion of our ventures, we need to follow back a Entity
to the particular Plan or work process that was utilized to create it.
This could be especially helpful while examining the im-pacts of various
work process setups or grasping the effect of explicit arranging choices
on the outcomes. Notwithstanding, PROV comes up short on worked in
system to guarantee the extremely durable and queryable relationship
among Entity and Plan examples. This limit has expected us to
investigate custom arrangements, for example, presenting extra metadata
or making custom connections among sub-stances and plans. While this
approach has worked by and by, it has added intricacy to our provenance
models and presented likely difficulties in keeping up with the
consistency of the connections over the long run.
\end{quote}

In spite of these difficulties, we have kept on working with the PROV
model and have fostered a few custom expansions and variations to
address the limits we have experienced. These changes have permitted us
to keep utilizing PROV successfully while keeping up with the
adaptability and versatility that are basic for our undertakings.

VI. CONCLUSIONS

\begin{quote}
Taking everything into account, Encompass has had the op-tion to
actually use the PROV structure to display provenance across numerous
frameworks and inside different IT areas. Our capacity to adjust the
PROV model to address the issues of our particular use cases has been a
critical calculate the outcome of our ventures. By utilizing the diagram
based design of PROV, we have had the option to catch definite
connections between substances, exercises, and specialists, guaranteeing
that we can follow the genealogy of information and figure out the work
processes that created it. This has furnished us with the adaptability
to address complex frameworks and work processes in an unmistakable and
effective way.

The utilization of PROV has empowered us to furnish our clients with the
certainty and straightforwardness they expect in the present information
driven world. By uncovering the provenance of individual outcomes, we
can assist our clients with understanding how information is produced
and handled, guaranteeing that they can believe the outcomes we give.
This is especially significant with regards to complex simulated
intelligence/ML applications, where the detectability and logic of
results are fundamental for guaranteeing the unwavering quality and
legitimacy of the models. Moreover, the capacity to follow and dissect
the exhibition of our frameworks through point by point provenance
information has permitted us to acquire significant experiences into the
adequacy of our work processes and distinguish amazing open doors for
develop-ment.

Quite possibly of the main illustration we have advanced during this
interaction is the significance of incorporating provenance following
into the center of our frameworks. By implanting provenance abilities
straightforwardly into our work processes, we have had the option to
catch the vital information without presenting pointless above or
intricacy. We have created both committed provenance apparatuses and
coordinated provenance highlights inside existing frameworks,
guaranteeing that provenance is consistently followed at each
\end{quote}

phase of the interaction. This approach has permitted us to accomplish
our targets without falling back on profoundly particular or excessively
complex executions of PROV.

Looking forward, we guess that our utilization of PROV will keep on
developing as our frameworks become more complicated and our necessities
become more particular. While we have not yet experienced a requirement
for exceptionally specific variants of PROV, we perceive that as our
tasks progress, we might have to investigate further developed
expansions or customizations to help new necessities. Later on, we might
have to additional improve the granularity or explicitness of the
provenance we catch, especially as we work with more mind boggling
simulated intelligence/ML models or as the size of our frameworks
develops.

In rundown, the utilization of PROV for provenance demon-strating has
been a significant device for Encompass, permit-ting us to catch and
track the heredity of information across different frameworks and work
processes. The adaptability, versatility, and extensibility of PROV have
pursued it an opti-mal decision for our activities, and we anticipate
proceeding to utilize and refine this model from here on out. Through
our continuous work with PROV, we are certain that we can meet the
developing necessities of our association and our clients, guaranteeing
that our frameworks stay straightforward, solid, and dependable.

\begin{quote}
REFERENCES
\end{quote}

{[}1{]} H. Knublauch and D. Kontokostas, ``Shapes Constraint Language
(SHACL),'' W3C RDF Data Shapes Working Group, W3C Recommen- dation,
2017. {[}Online{]}. Available: https://www.w3.org/TR/shacl/\\
{[}2{]} L. Moreau and P. Missier, ``PROV-DM: The PROV Data Model,''
World Wide Web Consortium, W3C Recommendation, 2013. {[}Online{]}.

Available: https://www.w3.org/TR/prov-dm/\\
{[}3{]} T. Lebo, S. Sahoo, and D. McGuinness, ``PROV-O: The PROV
Ontology,'' W3C Provenance Working Group, W3C Recommendation, 2013.
{[}Online{]}. Available: http://www.w3.org/TR/prov-o/\\
{[}4{]} Tom De Nies, Sara Magliacane, Ruben Verborgh, Sam Coppens, Paul
Groth, Erik Mannens, and Rik Van de Walle, ``Git2PROV: Exposing Version
Control System Content as W3C PROV,'' in \emph{Proceedings of} \emph{the
12th International Semantic Web Conference}, vol. II. Springer.

{[}Online{]}. Available: https://github.com/IDLabResearch/Git2PROV na,
E. Ogasawara, J. Dias, and {[}5{]} F. Costa, V. Silva, D. de Oliveira,
K. Oca˜\\
M. Mattoso, ``Capturing and querying workflow runtime provenance with
prov: A practical approach,'' in \emph{Proceedings of the Joint
EDBT/ICDT} \emph{2013 Workshops}, ser. EDBT '13. New York, NY, USA:
Association for Computing Machinery, 2013, p. 282--289.

{[}6{]} A. Prabhune, A. Zweig, R. Stotzka, M. Gertz, and J. Hesser,
``Prov2ONE: An Algorithm for Automatically Constructing ProvONE
Provenance Graphs,'' in \emph{Provenance and Annotation of Data and
Processes}, ser.

\begin{quote}
Lecture Notes in Computer Science, M. Mattoso and B. Glavic, Eds.\\
Cham: Springer International Publishing, 2016, pp. 204--208.
\end{quote}

{[}7{]} A. S. Butt and P. Fitch, ``A provenance model for control-flow
driven scientific workflows,'' \emph{Data \& Knowledge Engineering}, p.
101877, Feb. 2021. {[}Online{]}. Available:
https://linkinghub.elsevier.com/retrieve/ pii/S0169023X21000045\\
{[}8{]} N. J. Car, L. S. Stanford, and A. Sedgmen, ``Enabling Web
Service Request Citation by Provenance Information,'' in
\emph{Provenance and} \emph{Annotation of Data and Processes: 6th
International Provenance and} \emph{Annotation Workshop, IPAW 2016,
McLean, VA, USA, June 7-8,} \emph{2016, Proceedings}, M. Mattoso and B.
Glavic, Eds. Cham: Springer International Publishing, 2016, pp.
122--133. {[}Online{]}. Available:
http://dx.doi.org/10.1007/978-3-319-40593-3 10

\end{document}